\documentclass[conference,onecolumn,draft]{IEEEtran}
\usepackage[]{amsthm}
\usepackage{amsmath}
\usepackage{amssymb}
\usepackage[final]{graphicx}

\newcommand{\C}{\mathcal{C}}

\newcommand{\M}{\mathcal{M}}
\newcommand{\N}{\mathcal{N}}
\newcommand{\V}{\mathcal{V}}
\newcommand{\E}{\mathcal{E}}
\newcommand{\G}{\mathbf{G}}

\renewcommand{\deg}[1]{\mathsf{deg}\left(#1\right)}

\newcommand{\card}[1]{{|#1|}}

\newcommand{\field}{\mathbb{F}_q}
\newtheorem{prop}{Proposition}
\newtheorem{cor}{Corollary}
\newtheorem{thm}{Theorem}
\newtheorem{defn}{Definition}
\newtheorem{clry}{Corollary}
\newtheorem{lem}{Lemma}
\usepackage[draft=false]{hyperref}
\hypersetup{pdftitle={Coding with Constraints: Minimum Distance Bounds and Systematic Constructions},pdfkeywords={Constrained Coding; Distributed Storage; Systematic Codes; Error-Correcting Codes; Reed-Solomon Codes; Information Theory; Distributed Algorithms; Complexity; Finite Fields}}
\IEEEoverridecommandlockouts
\begin{document}
\title{Coding with Constraints: Minimum Distance Bounds and Systematic Constructions}
\author{\IEEEauthorblockN{Wael Halbawi, Matthew Thill \& Babak Hassibi}
\IEEEauthorblockA{Department of Electrical Engineering\\
California Institute of Technology\\
Pasadena, California 91125\\
Email: \{whalbawi,mthill,hassibi\}@caltech.edu\\
}
}
\maketitle
\begin{abstract}
We examine an error-correcting coding framework in which each coded symbol is constrained to be a function of a fixed subset of the message symbols. With an eye toward distributed storage applications, we seek to design systematic codes with good minimum distance that can be decoded efficiently. On this note, we provide theoretical bounds on the minimum distance of such a code based on the coded symbol constraints. We refine these bounds in the case where we demand a systematic linear code. Finally, we provide conditions under which each of these bounds can be achieved by choosing our code to be a subcode of a Reed-Solomon code, allowing for efficient decoding. This problem has been considered in multisource multicast network error correction. The problem setup is also reminiscent of locally repairable codes. 
\end{abstract}
\section{Introduction}
We consider a scenario in which we must encode $s$ message symbols using a length $n$ error-correcting code subject to a set of encoding constraints. Specifically, each coded symbol is a function of only a subset of the message symbols. This setup arises in various situations such as in the case of a sensor network in which each sensor can measure a certain subset of a set of parameters. The sensors would like to collectively encode the readings to allow for the possibility of measurement errors. Another scenario is one in which a client wishes to download data files from a set of servers, each of which stores information about a subset of the data files. The user should be able to recover all of the data even in the case when some of the file servers fail. Ideally, the user should also be able to download the files faster in the absence of server failures. To protect against errors, we would like the coded symbols to form an error-correcting code with reasonably high minimum distance. On the other hand, efficient download of data is permitted when the error-correcting code is of systematic form. Therefore, in this paper, we present an upper bound on the minimum distance of an error-correcting code when subjected to encoding constraints, reminiscent of the cut-set bounds presented in~\cite{Dikaliotis2011}. In certain cases, we provide a code construction that achieves this bound. Furthermore, we refine our bound in the case that we demand a systematic linear error-correcting code, and present a construction that achieves the bound. In both cases, the codes can be decoded efficiently due to the fact that our construction utilizes Reed-Solomon codes.
\subsection{Prior Work}
The problem of constructing error-correcting codes with constrained encoding has been addressed by a variety of authors. Dau et al.~\cite{Dau2013,Dau2014ISIT,Dau2014JSAC} considered the problem of finding linear MDS codes with constrained generator matrices. They have shown that, under certain assumptions, such codes exist over large enough finite fields, as well as over small fields in a special case.
A similar problem known as the weakly secure data exchange problem was studied in \cite{Yan2011},\cite{Yan2014}. The problem deals with a set of users, each with a subset of messages, who are interested in broadcasting their information securely when an eavesdropper is present. In particular, the authors of~\cite{Yan2014} conjecture the existence of secure codes based on Reed-Solomon codes and present a randomized algorithm to produce them.
The problem was also considered in the context of multisource multicast network coding in~\cite{Dikaliotis2011,Halbawi2014DRS,Halbawi2014DGC}. In~\cite{Halbawi2014DRS}, the capacity region of a simple multiple access network with three sources is achieved using Reed-Solomon codes. An analogous result is derived in~\cite{Halbawi2014DGC} for general multicast networks with 3 sources using Gabidulin codes.

There has been a recent line of work involving codes with local repairability properties, in which every parity symbol is a function of a predetermined set of data symbols \cite{Han2007, Huang2007, Gopalan2012, Pap2012, Tamo2013, Prakash2012, Kamath2013, Rawat2012,Tamo2014}. Another recent paper \cite{Mazumdar2014} represents code symbols as vertices of a partially connected graph. Each symbol is a function of its neighbors and, if erased, can be recovered from them. Our code also utilizes a graph structure, though only to describe the encoding procedure. There is not necessarily a notion of an individual code symbol being repairable from a designated local subset of the other code symbols.
\section{Problem Setup}
\begin{figure}
\centering
\includegraphics[scale=0.9]{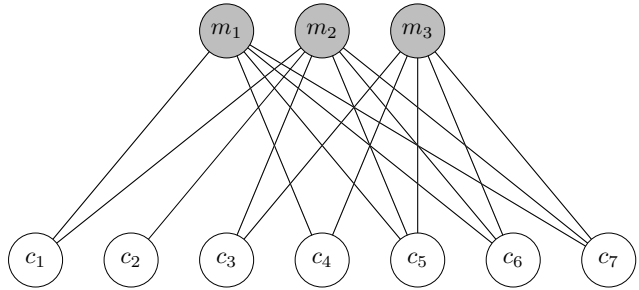}
\caption{A bipartite graph representing with 3 message symbols and 7 code symbols}
\label{fig:eg}
\end{figure}

Consider a bipartite graph $G = (\M,\V,\E)$ with $s =\card{\M} \leq \card{\V}=n$. The set $\E$ is the set of edges of the graph, with $(m_i, c_j) \in \E$ if and only if $m_i \in \M$ is connected to $c_j \in \V$. This graph defines a code where the vertices $\M$ correspond to message symbols and the vertices $\V$ correspond to codeword symbols. A bipartite graph with $s=3$ and $n=7$ is depicted in figure \ref{fig:eg}. Thus, if each $m_i$ and $c_j$ are assigned values in the finite field $\field$ with $q$ elements, then our messages are the vectors $\mathbf{m} = (m_1, \ldots, m_s) \in \field^s$ and our codewords are the vectors $\mathbf{c} = (c_1, \ldots, c_n) \in \field^n$. Each codeword symbol $c_j$ will be a function of the message symbols to which it is connected, as we will now formalize. 

Henceforth, $[\mathbf{c}]_\mathcal{I}$ is the subvector of $\mathbf{c}$ with elements indexed by $\mathcal{I} \subseteq \{1, ..., n\}$, and $[\mathbf{A}]_{i,j}$ is the $(i,j)^\text{th}$ element of a matrix $\mathbf{A}$. Let $\N(c_j)$ denote the neighborhood of $c_j \in \V$, i.e. $\N(c_j) = \{m_i \in \M : (m_i, c_j) \in \E\}$. Similarly, define $\N(m_i) = \{c_j : (m_i, c_j) \in \E\}$. We will also consider neighborhoods of subsets of the vertex sets, i.e. for $\V' \subseteq \V$, $\N(\V') = \cup_{c_j \in \V'} \N(c_j)$. The neighborhood of a subset of $\M$ is defined in a similar manner. Let $m_i$ take values in $\field$ and associate with each $c_j \in \V$ a function $f_j: \field^s \longrightarrow \field$. We restrict each $f_j$ to be a function of $\N(c_j)$ only. Now consider the set $\C = \{(c_1,\ldots,c_n):c_j = f_j(\mathbf{m}),\mathbf{m} \in \field^s\}$. The set $\C$ is an error-correcting code of length $n$ and size at most $q^s$. We will denote the minimum distance of $\C$ as $d(\C)$. If we restrict $f_j$ to be \emph{linear}, then we obtain a linear code with dimension at most $s$. 

The structure of the code's generator matrix can be deduced from the graph $G$. 
Let $\mathbf{g}_j \in \field^{s \times 1}$ be a column vector such that the $i^\text{th}$ entry is zero if $m_i \notin \N(c_j)$. Defining $f_j(\N(c_j)) = \mathbf{m}\mathbf{g}_j$ yields a linear function in which $c_j$ is a function of $\N(c_j)$ only, as required. A concatenation of the vectors $\mathbf{g}_j$ forms the following matrix:
\begin{equation}
\G = \left[\begin{array}{ccc}
| & & |\\
\mathbf{g}_1& \cdots & \mathbf{g}_n\\
| & & |
\end{array}\right]
\end{equation}
where $\G \in \field^{s \times n}$ is the generator matrix of the code $\C$. 

We associate with the bipartite graph $G = (\M,\V,\E)$ an adjacency matrix $\mathbf{A} \in \{0,1\}^{s \times n}$, where $[\mathbf{A}]_{i,j} = 1$ if and only if $(m_i, c_j) \in \E$. For the example in figure \ref{fig:eg}, this matrix is equal to
\begin{equation}
\mathbf{A} = \begin{bmatrix}
1 & 0 & 0 & 1 & 1 & 1 & 1\\
1 & 1 & 1 & 0 & 1 & 1 & 1\\
0 & 0 & 1 & 1 & 1 & 1 & 1\\
\end{bmatrix}
\label{eqn:adj}
\end{equation}
A \emph{valid}  generator matrix $\G$ (in generic form) is built from $\mathbf{A}$ by replacing non-zero entries with indeterminates. The choice of indeterminates (from a suitably-sized finite field $\field$) determines the dimension of the code and its minimum distance. For general linear codes, the Singleton bound (on minimum distance) is tight over large alphabets. 
In the presence of encoding constraints, the Singleton bound can be rather loose. In the next section, we derive an upper bound on the minimum distance of any code (linear or non-linear) associated with a bipartite graph. This bound is reminiscent of the cut-set bounds of Dikaliotis et al. in~\cite{Dikaliotis2011}.
\subsection{Subcodes of Reed-Solomon Codes}
\label{sec:RS}
Throughout this paper, we use the original definition of an $[n,k]_q$ Reed-Solomon code as in~\cite{Reed1960}, the $k$-dimensional subspace of $\mathbb{F}_q^n$ given by $\C_\text{RS} =\left\lbrace\left(m(\alpha_1), \ldots, m(\alpha_n)\right):\deg{m(x)} < k \right\rbrace$, where the $m(x)$ are polynomials over $\mathbb{F}_q$ of degree $\deg{m(x)}$, and the $\alpha_i \in \mathbb{F}_q$ are distinct (fixed) field elements. Each message vector $\mathbf{m} = \left(m_0, \ldots, m_{k-1}\right)$ is mapped to a message polynomial $m(x)=\sum_{i=0}^{k-1}m_i x^i$, which is then evaluated at the $n$ elements $\{\alpha_1,\alpha_2,\ldots,\alpha_n\}$ of $\mathbb{F}_q$, known as the defining set of the code. Reed-Solomon codes are MDS codes; their minimum distance attains the Singleton bound, i.e. $d(\C_\text{RS}) = n - k + 1$.

We can extract a subcode of a Reed-Solomon code that is valid for the bipartite graph $G = (\M, \V, \E)$ as follows: First, let $\field$ be a finite field with cardinality $q \geq n$. Associate to each $c_j \in \V$ a distinct element $\alpha_j \in \field$. Consider the $i^\text{th}$ row of the adjacency matrix $\mathbf{A}$ of $G$, and let $t_i(x) = \prod_{j: [\mathbf{A}]_{i,j} = 0}(x-\alpha_j)$. For example, ${t_3(x) = (x-\alpha_1)(x-\alpha_2)}$ corresponds to the the third row of $\mathbf{A}$ in \eqref{eqn:adj}. Choose $k$ such that $k > \deg{t_i(x)}$, $\forall i$. If $\mathbf{t}_i \in \field^k$ is the (row) vector of coefficients of $t_i(x)$ and $\G_{\text{RS}}$ is the generator matrix of a Reed-Solomon code with defining set $\{\alpha_1,\ldots,\alpha_n\}$ and dimension $k $, then $\mathbf{t}_i\mathbf{G}_\text{RS} = (t_i(\alpha_1), \ldots, t_i(\alpha_n))$ is a vector that is valid for the $i^\text{th}$ row of $\G$, i.e. if $[\mathbf{A}]_{i,j} = 0$ then $[\mathbf{t}_i\mathbf{G}_\text{RS}]_j = 0$. A horizontal stacking of the vectors $\mathbf{t}_i$ results in a transformation matrix $\mathbf{T}$ that will produce a valid generator matrix $\G$ from $\G_\text{RS}$:
\begin{equation}
\G = \mathbf{T}\G_\text{RS} = 
\begin{bmatrix}
&\mathbf{t}_1&\\
&\vdots&\\
&\mathbf{t}_s&\\
\end{bmatrix}
\begin{bmatrix}
1  & \cdots & 1\\
\alpha_1  & \cdots & \alpha_{n}\\
\vdots  & \ddots & \vdots\\
\alpha_1^{(k-1)}   & \cdots & \alpha_n^{(k-1)}
\end{bmatrix}
\end{equation}
The rank of $\G$ will be equal to the rank of $\mathbf{T}$, and the resulting code $\C$ will have a minimum distance $d(\C)$ that is determined by $\C_\text{RS}$. Indeed, $d(\C) \geq d(\C_\text{RS})$.
\section{Minimum Distance}
\begin{figure}
\centering
\includegraphics[scale=1]{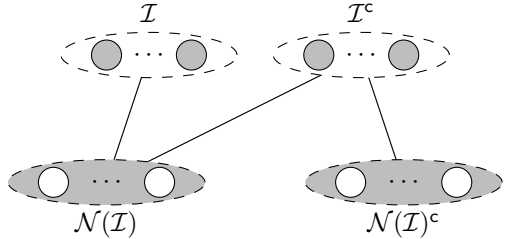}
\caption{Partitions of $\M$ and of $\V$ used in the proof of proposition \protect{\ref{prop:dmin}}. The set $\N(\mathcal{I})$ is a function of both $\mathcal{I}$ and $\mathcal{I}^\mathsf{c}$, while the set $\N(\mathcal{I})^\mathsf{c}$ is a function of $\mathcal{I}^\mathsf{c}$ only.}
\label{fig:dmin}
\end{figure}

In this section, an upper bound on the minimum distance of a code defined by a bipartite graph $G = (\M, \V, \E)$ is derived. The bound closely resembles the cut-set bounds of~\cite{Dikaliotis2011}. In most cases, this bound is tighter than the Singleton bound for a code of length $n$ and dimension $s$. For each $\M' \subseteq \M$ define $n_{\M'} :=\card{\N(\M')}$. This is the number of code symbols $c_j$ in $\V$ that are a function of the information symbols $\M'$. 
The following proposition characterizes the minimum distance of any code defined by $G$. 
\begin{prop}
\label{prop:dmin}
Fix a field $\field$. For any code $\C$ with $\card{\C} = q^s$ defined by a fixed graph $G = (\M,\V,\E)$, the minimum distance $d(\C)$ obeys
\begin{equation}
d(\C) \leq n_{\M'} - \card{\M'} + 1, \quad \forall \M' \subseteq \M. \label{eqn:dminbnd}
\end{equation}
\end{prop}
\begin{IEEEproof}
Working toward a contradiction, suppose $d(\C) > n_\mathcal{I} - \card{\mathcal{I}} + 1$ for some $\mathcal{I} \subseteq \M$. Let $\C'$ be the encoding of all message vectors $\mathbf{m}$ where $[\mathbf{m}]_{\mathcal{I}^\text{c}} \in \field^\card{\mathcal{I}^\text{c}}$ has some arbitrary but fixed value. Note that $[\mathbf{c}]_{\N(\mathcal{I})^\mathsf{c}}$ is the same for all $\mathbf{c} \in \C'$, since the symbols $\N(\mathcal{I})^\mathsf{c}$ are a function of $\mathcal{I}^\mathsf{c}$ only. Since $\card{\mathcal{I}} > n_\mathcal{I} - d(\C) + 1$, then by the pigeonhole principle there exist $\mathbf{c}_1, \mathbf{c}_2 \in \C'$ such that, without loss of generality, the first $n_\mathcal{I} - d(\C) + 1$ symbols of $[\mathbf{c}_1]_{\N(\mathcal{I})}$ and $[\mathbf{c}_2]_{\N(\mathcal{I})}$ are identical. Furthermore, $[\mathbf{c}_1]_{\N(\mathcal{I})^\mathsf{c}} = [\mathbf{c}_2]_{\N(\mathcal{I})^\mathsf{c}}$. Finally, since $\N(\mathcal{I})$ and $\N(\mathcal{I})^\text{c}$ partition $\V$, we obtain  $d_\text{H}(\mathbf{c}_1, \mathbf{c}_2) \leq {n - (n_\mathcal{I} - d(\C) + 1 + (n - n_\mathcal{I}))} = d(\C)-1$, a contradiction. Figure \ref{fig:dmin} illustrates the relation between $\mathcal{I}$ and the corresponding partition of $\V$.
\end{IEEEproof}
As a direct corollary, we obtain the following upper bound on $d(\C)$:
\begin{clry}
\begin{equation}
d(\C) \leq \min_{\M'\subseteq \M}\{n_{\M'} - \card{\M'}\} + 1 \label{eqn:dmin}
\end{equation}
\end{clry}

Our next task is to provide constructions of codes that achieve this bound.
\section{Systematic Construction}
In this section, we provide a code construction that achieves the minimum distance bound stated in corollary \ref{clry:dsys}. We appeal to Hall's Theorem, a well-known result in graph theory that establishes a necessary and sufficient condition for finding a matching in a bipartite graph. Some terminology needed from graph theory is defined in the following subsection.
\subsection{Graph Theory Preliminaries}
Let $G = (\mathcal{S},\mathcal{T},\E)$ be a bipartite graph. A \emph{matching} is a subset $\tilde{\E} \subseteq \E$ such that no two edges in $\tilde{\E}$ share a common vertex. A vertex is said to be \emph{covered} by $\tilde{\E}$ if it is incident to an edge in $\tilde{\E}$. An $\mathcal{S}$-\emph{covering} matching is one by which each vertex in $\mathcal{S}$ is covered. We will abuse terminology and say that an edge $e \in \tilde{\E}$ is \emph{unmatched} if $e \notin \tilde{\E}$. We can now state Hall's Theorem.
\begin{thm}
\label{thm:hall}
Let $G = (\mathcal{S},\mathcal{T},\E)$ be a bipartite graph. There exists an $\mathcal{S}$-covering matching if and only if $\card{\mathcal{S}'} \leq \N(\mathcal{S}')$ for all $\mathcal{S}' \subseteq \mathcal{S}$.
\end{thm}
For a proof of the theorem, see e.g. \cite[p.53]{LintWilsonBook}. 

Set $d_\mathsf{min} = \min_{\M'\subseteq \M}\{n_{\M'} - \card{\M'}\} + 1$. In order to construct a generator matrix $\G \in \field^{s\times n}$ for a code $\C$ with minimum distance $d_\mathsf{min}$, we will use an $[n,n-d_\mathsf{min} +1]$ Reed-Solomon code with generator matrix $\G_\text{RS}$. We will then extract $\mathcal{C}$ as a subcode using an appropriately built transformation matrix $\mathbf{T}$ to form $\G = \mathbf{T}\G_\text{RS}$ such that $\G$ is in systematic form, which implies that the dimension of $\C$ is $s$. Since $\C$ is a subcode of a code with minimum distance $d_\mathsf{min}$, we have $d(\C) \geq d_\mathsf{min}$. $\eqref{eqn:dmin}$ further implies that $d(\C) = d_\mathsf{min}$.

Our construction is as follows: consider a graph $G = (\M,\V,\E)$ defining $\C$, and define the set $\mathcal{A} = \{c_j : \N(c_j) = \M\}$, i.e. $\mathcal{A}$ is the set of code symbols that are a function of \textit{every} message symbol. Note that $\mathcal{A} \subseteq \N(\M')$ for every $\M' \subseteq \M$. Therefore, if $a = \card{\mathcal{A}}$ then the size of the neighborhood of $\N(\M')$ can be expressed as $n_{\M'} = r_{\M'} + a$, where $r_{\M'}$ is the cardinality of the set $\mathcal{R}(\M') = \N(\M') \setminus \mathcal{A}$. 
\begin{thm}
Let $G = (\M,\V,\E)$. Set $d_\mathsf{min} = \min_{\M'\subseteq \M}\{n_{\M'} - \card{\M'}\} + 1$ and $k_\mathsf{min} = n - d_\mathsf{min} + 1$. A linear code $\C$ with parameters $[n,s,d_\mathsf{min}]$ valid for $G$ can be constructed with a systematic-form generator matrix provided that $k_\mathsf{min} \geq r_\M$.
\end{thm}

\begin{IEEEproof}
First, we establish a bound on $a$. Note that since $n = n_\M = r_\M + a$ and $k_\mathsf{min} \geq r_\M$, then we have $a \geq d_\mathsf{min} - 1$.  Fix an arbitrary subset $\mathcal{A}^* \subseteq \mathcal{A}$ of size $a^* = a - (d_\mathsf{min} - 1)$, which is guaranteed to exist by virtue of the bound on $a$, and let $\mathcal{B} = \mathcal{A}\setminus\mathcal{A}^*$. Now, we focus on a particular subgraph of $G$ defined by $G^* = (\M,\V^*,\E^*)$ where $\V^* = \V \setminus \mathcal{B}$, and $\E^*= \{(m_i,c_j) \in \E: c_j \in \V^*\}$ is the edge set corresponding to this subgraph.
Since $n_{\M'} = r_{\M'} + a$, then from the definition of $d_\mathsf{min}$ we have 
\begin{equation}
\card{\M'} \leq r_{\M'} + a - (d_\mathsf{min} - 1), \quad \forall \M' \subseteq \M \label{eqn:Hall1}
\end{equation}
The neighborhood of every subset $\M'$ when restricted to $\V^*$ is exactly $\N^*(\M') = \mathcal{R}(\M') \cup \mathcal{A}^*$, with cardinality $n^*_{\M'} = r_{\M'} + a^*$. The bounds \eqref{eqn:Hall1} can now be expressed in a way suitable for the condition of Hall's theorem: 
\begin{equation}
\card{\M'} \leq n^*_{\M'}, \quad \forall \M' \subseteq \M \label{eqn:Hall2}
\end{equation}
An $\mathcal{M}$-covering matching in $G^*$ can be found by letting $\mathcal{S} = \M$ and $\mathcal{T} = \V^*$ in theorem~\ref{thm:hall}. Let $\tilde{\E} = \{(m_i,c_{{j}(i)})\}_{i=1}^s\subseteq \E^*$ be such a matching, and $\tilde{\V}$ the subset of $\V^*$ that is covered by $\tilde{\E}$. Let $\mathbf{A}_{\tilde{\E}}$ be the adjacency matrix of $G$ when the edge set $\{(m_i,c_j) \in \E: c_j \in \tilde{\V}, j \neq j(i)\}$ is removed. The number of zeros in any row of $\mathbf{A}_{\tilde{\E}}$ is at most $n - d_\mathsf{min}$. To see this, note that the edges in $\E$ incident to $\mathcal{B}$ are not removed by the matching, and every $m_i \in \mathcal{M}$ is connected to at least one vertex in $\V^*$.
Next, we build a valid $\G$ for $G$ using $\mathbf{A}_{\tilde{\E}}$, utilizing the method described in section \ref{sec:RS}. Fix a $[n,n-d_\mathsf{min}+1]$ Reed-Solomon code with generator matrix $\G_\text{RS}$ and defining set $\{\alpha_1, \ldots, \alpha_n\}$. The $i^\text{th}$ transformation polynomial is $t_i(x)=\prod_{j:[\mathbf{A}_{\tilde{\E}}]_{i,j}=0}(x-\alpha_i)$. Since the number of zeros in any row of $\mathbf{A}_{\tilde{\E}}$ is at most $n - d_\mathsf{min}$, we have $\deg {t_i(x)} \leq n - d_\mathsf{min} = k - 1$ for all $i$. We use the $t_i(x)$, after normalizing by $t_i(\alpha_{j(i)})$, to construct a transformation matrix $\mathbf{T}$ and then $\G = \mathbf{T}\G_\text{RS}$ is valid for $G$. Note that  $\G$ is in systematic form due the fact that the columns of $\mathbf{A}_{\tilde{\E}}$ indexed by $\{j(i)\}_{i=1}^s$ form a permutation of the identity matrix of size $s$. Lastly, $d(\C) = d_\text{min}$ since $d(\C) \leq d_\mathsf{min}$ by corollary \eqref{eqn:dmin}, and $d(\C) \geq d_\mathsf{min}$ since $\C$ is a subcode of a code with minimum distance $d_\mathsf{min}$.
\end{IEEEproof}
\section{Minimum Distance for Systematic Linear Codes}
In this section, we will restrict our attention to the case where a code  valid for $G$ is linear, so that each $c_j \in \V$ is a linear function of the message symbols $m_i \in \N(c_j)$. We seek to answer the following: What is the greatest minimum distance attainable by a \textit{systematic} linear code valid for $G$? 

Any systematic code must correspond to a matching $\tilde{\E} \subseteq \E$ which identifies each message symbol $m_i \in \M$ with a unique codeword symbol $c_{j(i)} \in \V$, where $j(i) \in \{1, \ldots, n\}$. Explicitly, $\tilde{\E}$ consists of $s$ edges of the form $\{(m_i, c_{j(i)})\}$ for $i = 1, \ldots, s$ such that $c_{j(i_1)} \ne c_{j(i_2)}$ for $i_1 \ne i_2$.  As before, $\tilde{\V}$ is the subset of vertices in $\V$ which are involved in the matching: $\tilde{\V} = \{c_{j(i)}\}_{i = 1}^s$. Our code becomes systematic by setting $c_{j(i)} = m_i$ for $i = 1, \ldots, s$, and choosing each remaining codeword symbol $c_j \notin \tilde{\V}$ to be some linear function of its neighboring message symbols $m_i \in \N(c_j)$.

\begin{defn} 
For $G = (\M, \V, \E)$, let $\tilde{\E} \subseteq \E$ be an $\M$-covering matching so that $\tilde{\E} = \{(m_i, c_{j(i)})\}_{i = 1}^s$. Let $\tilde{\V} = \{c_{j(i)}\}_{i = 1}^s$ be the vertices in $\V$ which are covered by $\tilde{\E}$. Define the \textit{matched adjacency matrix} $\mathbf{A}_{\tilde{\E}} \in \{0, 1\}^{s \times n}$ so that $[\mathbf{A}_{\tilde{\E}}]_{i, j} = 1$ if and only if either $(m_i, c_j) \in \tilde{\E}$, or $c_j \notin \tilde{\V}$ and $(m_i, c_j) \in \E$. In other words, $\mathbf{A}_{\tilde{\E}}$ is the adjacency matrix of the bipartite graph formed by starting with $G$ and deleting the edges $\{(m_i, c_j) \in \E ~:~ c_j \in \tilde{\V} \text{ and } j \ne j(i)\}$. 
\end{defn}

\begin{defn} 
Let $\tilde{\E} \subseteq \E$ be a matching for the $G = (\M, \V, \E)$ which covers $\M$. Let $z_{\tilde{\E}}$ be the maximum number of zeros in any row of the corresponding matched adjacency matrix $\mathbf{A}_{\tilde{\E}}$, and define $k_{\tilde{\E}} := z_{\tilde{\E}} + 1$.  Furthermore, define $k_\mathsf{sys} = \text{min}_{\tilde{\E}} k_{\tilde{\E}}$ where $\tilde{\E}$ ranges over all matchings for $G$ which cover $\M$, and $d_\mathsf{sys} = n - k_\mathsf{sys} + 1$. 
\end{defn} 

\begin{lem} 
For a given bipartite graph $G = (\M, \V, \E)$ which merits a matching that covers $\M$, we have 
\begin{equation} 
s \le k_\mathsf{min} \le k_\mathsf{sys} \le n
\end{equation} 
and 
\begin{equation} 
d_\mathsf{sys} \le d_\mathsf{min}. 
\end{equation} 
\end{lem} 

\begin{IEEEproof} 
Let $\mathbf{A}$ be the adjacency matrix of $G$. 

For any subset $\M' \subseteq \M$ we have $d_\mathsf{min} \le n_{\M'} - |\M'| + 1$, and likewise $k_\mathsf{min} = n - d_\mathsf{min} + 1 \ge |\M'| + (n-n_{\M'})$. Taking $\M' = \M$ (and noting that in our framework, every $c_j \in \V$ is connected to at least one vertex in $\M$, hence $n_{\M} = n$) we obtain $k_\mathsf{min} \ge s$. 

Now choose a set $\M'$ for which the above relation holds with equality, that is, $k_\mathsf{min} = |\M'| + (n-n_{\M'})$. Since $\N(\M')$ is simply the union of the support sets of the rows of $\mathbf{A}$ corresponding to $\M'$, then each of these rows must have at least $n - n_{\M'} = |\N(\M')^\mathsf{c}|$ zeros. Furthermore, any matching $\tilde{\E}$ which covers $\M$ must identify the rows of $\M'$ with columns of $\N(\M')$. Thus, in the matched adjacency matrix $\mathbf{A}_{\tilde{\E}}$, the row corresponding to $j \in \M'$ must have $|\M'| - 1$ zeros in the columns of $\N(\M)$ which are matched to $\M' \setminus \{j\}$, in addition to the $n - n_{\M'}$ zeros in the columns corresponding to $\N(\M')$. This gives us $k_{\tilde{\E}} \ge |\M'| + (n-n_{\M'})$ for each matching $\tilde{\E}$, hence $k_\mathsf{sys} \ge k_\mathsf{min}$. It follows directly that $d_\mathsf{sys} \le d_\mathsf{min}$. Finally, it is clear from definition that for any $\M$-covering matching $\tilde{\E}$ we must have that $k_{\tilde{\E}}$ is less than the length of the adjacency matrix $\mathbf{A}$, which is $n$, hence $k_\mathsf{sys} \le n$. 
\end{IEEEproof}
\begin{cor}
\label{clry:dsys}
Let $G = (\M, \V, \E)$ be a bipartite graph which merits a systematic linear code. The largest minimum distance obtainable by a systematic linear code is $d_\mathsf{sys}$. 
\end{cor} 
\begin{IEEEproof} 
Let $\C$ be a systematic linear code which is valid for $G$. Then $\C$ must have a codeword containing at least $k_\mathsf{sys} - 1$ zeros, i.e. a codeword of Hamming weight at most $n - k_\mathsf{sys} + 1 = d_\mathsf{sys}$. Since the code is linear, this Hamming weight is an upper bound for its minimum distance, so $d(\C) \le d_\mathsf{sys}$. 

It remains to see that there are systematic linear codes which are valid for $G$ and achieve a minimum distance of $d_\mathsf{sys}$. Let $\tilde{\E}$ be an $\M$-covering matching for $G$ such that $k_{\tilde{\E}} = k_\mathsf{sys}$. Then for any $k \ge k_\mathsf{sys}$, we claim that an $[n, k]$ Reed-Solomon code contains a systematic linear subcode that is valid for $G$. Indeed, choose a set of $n$ distinct elements $\{\alpha_i\}_{i = 1}^n \subseteq \field$ as the defining set of our Reed-Solomon code. Then to form our subcode's generator matrix $\mathbf{G}$, note that (as mentioned before) $\mathbf{G}$ must have zero entries in the same positions as the zero entries of $\mathbf{A}_{\tilde{\E}}$, and indeterminate elements in the remaining positions. There are at most $k_\mathsf{sys} - 1$ zeros in any row of $\mathbf{A}_{\tilde{\E}}$ (and at \textit{least} $s-1$ zeros in each row, since there must be $s$ columns which have nonzero entries in exactly one row). For each row $i \in \{1,\ldots,s\}$ of $\mathbf{A}_{\tilde{\E}}$, let $\mathcal{I}_i \subseteq \{1,\ldots,n\}$ be the set of column indices $j$ such that $[\mathbf{A}_{\tilde{\E}}]_{i, j}  = 0$. Then form the polynomial $t_i(x) = \prod_{j \in \mathcal{I}_i} (x - \alpha_j)$ and normalize by $t_i(\alpha_{j(i)})$, which accordingly has degree at most $k_\mathsf{sys}$ (and at least $s-1$). We now set the $i^{th}$ row of $\mathbf{G}$ to be $(t_i(\alpha_1),\ldots, t_i(\alpha_n))$, and we see that by construction this row has zeros precisely at the indices $j \in \mathcal{I}_i$ as desired. 

The rows of $\mathbf{G}$ generate a code with minimum distance at least that of the original Reed-Solomon code, which is $n - k + 1$. Furthermore, by setting $k = k_\mathsf{sys}$ for our Reed-Solomon code, we see this new code $\C$ has minimum distance at least $n - k_\mathsf{sys} + 1 = d_\mathsf{sys}$. Since by our previous argument, $d(\C) \le d_\mathsf{sys}$, the minimum distance of $\C$ must achieve $d_\mathsf{sys}$ with equality. 
\end{IEEEproof} 
\section{Achievability Using MDS Codes}
Throughout this paper, we have utilized Reed-Solomon codes to construct systematic linear codes valid for a particular $G=(\M,\V,\E)$ that attain the highest possible distance. It is worth mentioning that this choice is not necessary and in fact, the Reed-Solomon code utilized can be replaced with any linear MDS code with the same parameters.
\begin{lem}
Fix an arbitrary $[n,k]$ linear MDS code $\C$. For any $\mathcal{I} \subseteq [n]$ where $\card{\mathcal{I}} \leq k-1$ , there exists $\mathbf{c} \in \C$ such that $[\mathbf{c}]_\mathcal{I} = \mathbf{0}$.
\begin{IEEEproof}
Let $\G = [\mathbf{g}_i]_{i=1}^n$ be the generator matrix of $\C$ and let $\G_\mathcal{I}=[\mathbf{g}_i]_{i \in \mathcal{I}}$. Since $\card{\mathcal{I}} \leq k-1$, $\G_\mathcal{I}$ has full column rank and so it has a non-trivial left nullspace of dimension $k - \card{\mathcal{I}}$. If $\mathbf{h}$ is any vector in that nullspace then $\mathbf{c} = \mathbf{h}\G$ is such that $[\mathbf{c}]_\mathcal{I} = \mathbf{0}$.
\end{IEEEproof}
\end{lem}
Therefore, to produce a valid linear code $\C$ for $G=(\M,\V,\E)$ with $d(\C) = d^*$, where $d^* \leq n_{m_i}$ for all $m_i \in \M$, we fix an arbitrary $[n,n-d^*+1]$ MDS code and then select vectors ${\mathbf{h}_1,\ldots,\mathbf{h}_s}$ such that $\mathbf{h}_i$ is in the left nullspace of $\G_{\mathcal{I}_i}$, where $\mathcal{I}_i =\{j:\mathbf{A}_{i,j} = 0\}$. Note that the specific selection of the $\mathbf{h}_i$ determines the dimension of $\C$.
For a systematic construction, in which the dimension of the code is guaranteed to be $s$, some extra care has to be taken when choosing the $\mathbf{h}_i$. We must choose each $\mathbf{h}_i$ such that its not in the nullspace of $\mathbf{g}_{j(i)}$, which the column corresponding to the systematic coordinate $c_{j(i)}$.
\section{Example}
In this section, we construct a systematic linear code that is valid for the graph in figure \ref{fig:eg}. The bound of corollary \ref{eqn:dmin} asserts that $d(\C) \leq 5$ for any $\C$ valid for $G$. However,  lemma \ref{clry:dsys} shows that $d(\C_\mathsf{sys}) \leq4$ for any valid systematic linear code $\C_\mathsf{sys}$. A matching achieving this bound is given by the edges $\tilde{\E} = \{(m_1,v_1),(m_2,v_2),(m_3,v_3)\}$ and so the edges removed from the graph are $\{(m_2,v_1),(m_2,v_3)\}$. The new adjacency matrix $\mathbf{A}_{\tilde{\E}}$ is given by,
\begin{equation}
\mathbf{A}_{\tilde{\E}} = \begin{bmatrix}
1 & 0 & 0 & 1 & 1 & 1 & 1\\
\mathbf{0} & 1 & 0 & \mathbf{0} & 1 & 1 & 1\\
0 & 0 & 1 & 1 & 1 & 1 & 1\\
\end{bmatrix}
\end{equation}
where boldface zeros refer to those edges removed from $G$ because of the matching $\tilde{\E}$.

A generator matrix which is valid for $\mathbf{A}_{\tilde{\E}}$ can be constructed from that of a $[7,4]$ Reed-Solomon code over $\mathbb{F}_7$ with defining set $\{0,1,\alpha,\ldots,\alpha^5\}$ where $\alpha$ is a primitive element in $\mathbb{F}_7$, using the method described in \ref{sec:RS}. 

The polynomials corresponding to the transformation matrix are given by,
\begin{IEEEeqnarray}{rCl}
t_1(x) & = & \alpha^{5}(x-1)(x-\alpha) \\
t_2(x) & = & \alpha^{4}x(x-\alpha)(x-\alpha^2) \\
t_3(x) & = &\alpha^{3}x(x-1)
\end{IEEEeqnarray}
Finally, the systematic generator matrix for $\C_\mathsf{sys}$ is,
\begin{equation}
\G_\mathsf{sys}= \begin{bmatrix}
1 & 0 & 0 & \alpha^2 & \alpha^5 & 1 & \alpha^5 \\
0 & 1 & 0 & 0 & 1 & \alpha^4 & 1 \\
0 & 0 & 1 & \alpha^5 & \alpha^5 & \alpha^2 & 1
\end{bmatrix}
\end{equation}
\section{Conclusion}
In this paper, we have studied the problem of analyzing and designing error-correcting codes when the encoding of every coded symbol is restricted to a subset of the message symbols. We obtain an upper bound on the minimum distance of any such code, similar to the cut-set bounds of~\cite{Dikaliotis2011}. By providing an explicit construction, we show that under certain assumptions this bound is achievable. Furthermore, the field size required for the construction scales linearly with the code length. The second bound is on the minimum distance of linear codes with encoding constraints when the generator matrix is required to be in systematic form. We provide a construction that always achieves this bound. Since all of our constructions are built as subcodes of Reed-Solomon codes, they can be decoded efficiently using standard Reed-Solomon decoders. For future work, it remains to show that the first upper bound is achievable in general over small fields.
\bibliographystyle{IEEEtran}
\bibliography{IEEEabrv,library}

\begin{thebibliography}{10}
\providecommand{\url}[1]{#1}
\csname url@samestyle\endcsname
\providecommand{\newblock}{\relax}
\providecommand{\bibinfo}[2]{#2}
\providecommand{\BIBentrySTDinterwordspacing}{\spaceskip=0pt\relax}
\providecommand{\BIBentryALTinterwordstretchfactor}{4}
\providecommand{\BIBentryALTinterwordspacing}{\spaceskip=\fontdimen2\font plus
\BIBentryALTinterwordstretchfactor\fontdimen3\font minus
  \fontdimen4\font\relax}
\providecommand{\BIBforeignlanguage}[2]{{%
\expandafter\ifx\csname l@#1\endcsname\relax
\typeout{** WARNING: IEEEtran.bst: No hyphenation pattern has been}%
\typeout{** loaded for the language `#1'. Using the pattern for}%
\typeout{** the default language instead.}%
\else
\language=\csname l@#1\endcsname
\fi
#2}}
\providecommand{\BIBdecl}{\relax}
\BIBdecl

\bibitem{Dikaliotis2011}
T.~K. Dikaliotis, T.~Ho, S.~Jaggi, S.~Vyetrenko, H.~Yao, M.~Effros, J.~Kliewer,
  and E.~Erez, ``\BIBforeignlanguage{English}{{Multiple-Access Network
  Information-Flow and Correction Codes}},''
  \emph{\BIBforeignlanguage{English}{IEEE Trans. Inf. Theory}}, vol.~57, no.~2,
  pp. 1067--1079, Feb. 2011.

\bibitem{Dau2013}
S.~H. Dau, W.~Song, Z.~Dong, and C.~Yuen, ``{Balanced Sparsest generator
  matrices for MDS codes},'' in \emph{Inf. Theory Proc. (ISIT), 2013 IEEE Int.
  Symp.}, 2013, pp. 1889--1893.

\bibitem{Dau2014ISIT}
S.~H. Dau, W.~Song, and C.~Yuen, ``{On the existence of MDS codes over small
  fields with constrained generator matrices},'' in \emph{Inf. Theory (ISIT),
  2014 IEEE Int. Symp.}, Jun. 2014, pp. 1787--1791.

\bibitem{Dau2014JSAC}
------, ``{On Simple Multiple Access Networks},'' \emph{IEEE J. Sel. Areas
  Commun.}, vol. 8716, no. 0733, pp. 1--1, 2014.

\bibitem{Yan2011}
M.~Yan and A.~Sprintson, ``{Weakly Secure Network Coding for Wireless
  Cooperative Data Exchange},'' in \emph{Glob. Telecommun. Conf. (GLOBECOM
  2011), 2011 IEEE}, Dec. 2011, pp. 1--5.

\bibitem{Yan2014}
M.~Yan, A.~Sprintson, and I.~Zelenko, ``{Weakly Secure Data Exchange with
  Generalized Reed Solomon Codes},'' 2014, pp. 1366--1370.

\bibitem{Halbawi2014DRS}
W.~Halbawi, T.~Ho, H.~Yao, and I.~Duursma, ``{Distributed reed-solomon codes
  for simple multiple access networks},'' in \emph{2014 IEEE Int. Symp. Inf.
  Theory}.\hskip 1em plus 0.5em minus 0.4em\relax IEEE, Jun. 2014, pp.
  651--655.

\bibitem{Halbawi2014DGC}
W.~Halbawi, T.~Ho, and I.~Duursma, ``{Distributed gabidulin codes for
  multiple-source network error correction},'' in \emph{2014 Int. Symp. Netw.
  Coding}.\hskip 1em plus 0.5em minus 0.4em\relax IEEE, Jun. 2014, pp. 1--6.

\bibitem{Han2007}
J.~Han and L.~A. Lastras-Montano, ``{Reliable Memories with Subline
  Accesses},'' in \emph{2007 IEEE Int. Symp. Inf. Theory}.\hskip 1em plus 0.5em
  minus 0.4em\relax IEEE, Jun. 2007, pp. 2531--2535.

\bibitem{Huang2007}
C.~Huang, M.~Chen, and J.~Li, ``{Pyramid Codes: Flexible Schemes to Trade Space
  for Access Efficiency in Reliable Data Storage Systems},'' in \emph{Sixth
  IEEE Int. Symp. Netw. Comput. Appl. (NCA 2007)}.\hskip 1em plus 0.5em minus
  0.4em\relax IEEE, Jul. 2007, pp. 79--86.

\bibitem{Gopalan2012}
P.~Gopalan, C.~Huang, H.~Simitci, and S.~Yekhanin, ``{On the Locality of
  Codeword Symbols},'' \emph{IEEE Trans. Inf. Theory}, vol.~58, no.~11, pp.
  6925--6934, Nov. 2012.

\bibitem{Pap2012}
D.~S. Papailiopoulos and A.~G. Dimakis, ``{Locally repairable codes},'' in
  \emph{2012 IEEE Int. Symp. Inf. Theory Proc.}\hskip 1em plus 0.5em minus
  0.4em\relax IEEE, Jul. 2012, pp. 2771--2775.

\bibitem{Tamo2013}
I.~Tamo, D.~S. Papailiopoulos, and A.~G. Dimakis, ``{Optimal locally repairable
  codes and connections to matroid theory},'' in \emph{2013 IEEE Int. Symp.
  Inf. Theory}.\hskip 1em plus 0.5em minus 0.4em\relax IEEE, Jul. 2013, pp.
  1814--1818.

\bibitem{Prakash2012}
N.~Prakash, G.~M. Kamath, V.~Lalitha, and P.~V. Kumar, ``{Optimal linear codes
  with a local-error-correction property},'' in \emph{2012 IEEE Int. Symp. Inf.
  Theory Proc.}\hskip 1em plus 0.5em minus 0.4em\relax IEEE, Jul. 2012, pp.
  2776--2780.

\bibitem{Kamath2013}
G.~M. Kamath, N.~Prakash, V.~Lalitha, and P.~V. Kumar, ``{Codes with local
  regeneration},'' in \emph{2013 Inf. Theory Appl. Work.}\hskip 1em plus 0.5em
  minus 0.4em\relax IEEE, Feb. 2013, pp. 1--5.

\bibitem{Rawat2012}
A.~S. Rawat, O.~O. Koyluoglu, N.~Silberstein, and S.~Vishwanath, ``{Optimal
  Locally Repairable and Secure Codes for Distributed Storage Systems},'' Oct.
  2012.

\bibitem{Tamo2014}
I.~Tamo and A.~Barg, ``A family of optimal locally recoverable codes,''
  \emph{Information Theory, IEEE Transactions on}, vol.~60, no.~8, pp.
  4661--4676, Aug 2014.

\bibitem{Mazumdar2014}
A.~Mazumdar, ``{Storage Capacity of Repairable Networks},''
  \emph{arXiv:1408.4862}, Aug. 2014.

\bibitem{Reed1960}
I.~Reed and G.~Solomon, ``{Polynomial codes over certain finite fields},''
  \emph{J. Soc. Ind. Appl. Math.}, 1960.

\bibitem{LintWilsonBook}
J.~H. {Van Lint} and R.~M. Wilson, \emph{{A Course in Combinatorics}}.\hskip
  1em plus 0.5em minus 0.4em\relax Cambridge University Press, 2011.

\end{thebibliography}
\end{document}